# Identifying Relevant Eigenimages – a Random Matrix Approach

Yu Ding, Yiu-Cho Chung, Kun Huang, and Orlando P. Simonetti

*Abstract*—**Dimensional reduction of high dimensional data can be achieved by keeping only the relevant eigenmodes after principal component analysis. However, differentiating relevant eigenmodes from the random noise eigenmodes is problematic. A new method based on the random matrix theory and a statistical goodness-of-fit test is proposed in this paper. It is validated by numerical simulations and applied to real-time magnetic resonance cardiac cine images.**

## I. Introduction

In the fields of pattern recognition, machine learning, and computer vision, eigen-decomposition based methods such as principal component analysis (PCA) [1], multidimensional scaling (MDS) [2], and spectral clustering [3] have been widely adopted for the purpose of dimensionality reduction, data representation, and clustering. A common fundamental issue in these approaches is to determine how many eigenvectors are needed for further analysis (e.g., data representation and clustering). Traditionally, the choice of the number of principal eigenvectors is made in an empirical or ad hoc manner (e.g., the knee point of the eigenvalue plot). In PCA, one of the goals is to find a low-dimensional representation of the data so that the residuals in the ambient space are randomly distributed.

Recent development of the random matrix theory (RMT) [4] brings a new tool to this old problem. RMT has been successfully applied to study nuclear energy level [4], quantum chaos [5], EEG data [6], communication systems [7], and the financial market [8, 9]. Usually, a data matrix is the sum of a low-rank signal matrix and a full-rank noise matrix. Sengupta et al [10] studied the probability

Y. Ding is with The Ohio Sate University, Columbus, OH 43210 USA (phone: 614-293-2710; fax: 614-293-2561; e-mail: Yu.Ding@ osumc.edu).
Y.-C Chung, is with Siemens Healthcare, Inc., Columbus, OH 43210 USA. (e-mail: yiucho.chung@siemens.com).

distribution function (PDF) of singular values of such data matrices, and found that part of the small singular values follows the distribution of Marcenko-Pastur (MP) law [11]. One of the important results from RMT is that the empirical PDF of eigenvalues of the random variable co-variance matrix follows the MP-law. In other words, by removing a proper number of eigenmodes from a data matrix, the residual matrix behaves like a random matrix.

We propose a new method to identify the number of relevant eigenmodes of a data matrix by maximizing the similarities between the residual data matrix and a random matrix. The kernel of this method is a parameter estimation to maximize the statistical goodness-of-fit (GOF) [12] between the MP-law and a subset of the eigenvalues. We refer to this approach as the RMT-fitting method. When a data matrix contains independent and identically-distributed (IID) noise, the RMT-fitting method has only one parameter to estimate, the number of relevant eigenvalues. When non-IID noise appears, satisfactory GOF can be achieved by varying another parameter.

In this study, we validate the RMT-fitting method using numerical simulation. This technique is then demonstrated in the analysis of real-time magnetic resonance (MR) cardiac images acquired by parallel imaging techniques [13, 14] which introduce non-IID noise. We find that the RMT-fitting method successfully identifies the number of relevant eigenimages in this medical imaging application.

## II. Theory

We focus on a data set that can be represented by a 2-D matrix, the first dimension representing the temporal sampling, and the second dimension representing the spatial sampling, such as a movie taken from a CCD camera. For an $M \times N$ data matrix $\boldsymbol{A} = \boldsymbol{S_o} + \boldsymbol{n}$ (with no loss of generality, assuming $M<N$), where matrix $\boldsymbol{S_o}$ has a low-rank structure and entries of matrix $\boldsymbol{n}$ are IID noise, the $M \times M$ temporal covariance matrix (TCM) is given by $\boldsymbol{A}^T\boldsymbol{A}/N$. $\boldsymbol{S_o}$ represents the underlying signal such as a noiseless dynamic cardiac image series with $M$ frames, each frame containing $N$ pixels. If $\boldsymbol{S_o}$ has $r$ ($r < M$) independent signal modes with large variances when comparing to the noise variance[15], then the PDF of the smallest $M$-$r$ TCM eigenvalues follows the MP-law [10]:

$$p(\lambda) = \frac{\alpha}{2\pi\sigma^2\lambda}\sqrt{\max(0,(\lambda_{\max}-\lambda)(\lambda-\lambda_{\min}))} \qquad (1)$$

where $\alpha = (M-r)/N$ , $\lambda_{\max,\min} = \sigma^2(1 \pm \sqrt{\alpha})$, and $\sigma^2$ is the mean noise variance. When no signal appears in matrix $\boldsymbol{A}$, $r$ = 0 in Eq.(1). Throughout the paper, all eigenvalues mentioned are of TCM, and sorted in a descending order.

When non-IID noise appears, the TCM eigenvalue PDF is unknown. However, we observed that the MP-law is still

K. Huang is with The Ohio Sate University, Columbus, OH 43210 USA (e-mail: khuang@bmi.osu.edu).
O. P. Simonetti is with The Ohio Sate University, Columbus, OH 43210 USA (e-mail: Orlando.Simonetti@osumc.edu).

applicable when a data matrix is corrupted by two types of non-IID noise, spatially correlated Gaussian noise and spatially variant Gaussian noise. Widely used signal post-processing algorithms such as interpolation or low-pass filtering may cause spatial correlation of noise. The recently developed MRI fast imaging method called 'parallel imaging' can introduce spatially variant noise. We found that by optimizing two parameters, $r$ and $N$ of Eq. (1), the MP-law can fit reasonably well to the PDF of the smallest eigenvalues of the TCM of a dynamic MRI cardiac cine image series acquired using parallel imaging on a commercial MRI scanner.

The RMT-fitting method searches the maxima of the statistical goodness-of-fit (GOF) in a 2-D parameter space (the $r$, $N$-plane). Parameter $\sigma^2$ is the mean of last $M$-$r$ eigenvalues, which is a function of $r$. We use the Kolmogorov-Smirnov (KS) test or the Cramer-von-Mises (CVM) test [16] as the GOF function. We test the effectiveness of the RMT-fitting method using numerical simulations, and then apply it to in-vivo real-time MR cardiac cine images to identify the relevant eigenimages.

## III. Method

All the images for simulation and analysis were acquired in healthy volunteers. This study is approved by our institution's Human Subjects Committee. All subjects gave written informed consent to participate in these studies.

### A. Numerical Simulation

A segmented cardiac MR cine [17, 18] series consisting of 20 frames, evenly distributed over a single cardiac cycle, was acquired in a healthy volunteer. The in-plane spatial resolution, temporal resolution, and image matrix were 1.4 mm, 47.7 ms, and 256 × 208 respectively ($M = 20$, $N = 256 \times 208$). $\lambda_{max} = 2.65 \times 10^5$ and $\lambda_{min} = 72$ are the largest and the smallest eigenvalues. A dynamic cardiac cine series over more than one cardiac cycle can be simulated by repeating the original series β times (β>1). Regardless of β, the data matrix ($M = 20 \times \beta$, $N = 256 \times 208$) has a fixed rank = 20.

The first simulation was designed to explore the performance of the RMT-fitting method as a function of the additive noise variance. An image series with 4 repetitions (β =4) was constructed. Three different types of noise were added to it: (i) zero-mean Gaussian IID noise; (ii) spatial correlated Gaussian noise, *e.g.* the noise in (i) filtered by a 2D spatial low-pass filtering (LPF), with a 2-D square box-car function keeping 25% of the Fourier space and (iii) spatially varying Gaussian noise generated by multiplying uniform Gaussian noise by a spatially variant template formed by uncorrelated random numbers following a uniform distribution between zero and one (the same template was used in all 80 frames). In each case, the noise variance was varied from $\lambda_{min}$ to $\lambda_{max}$. We defined the critical noise variance $V_c$ as the noise variance at which the RMT-fitting method find 19 relevant eigenimages. We also varied the LPF cutoff to check the accuracy of the fitted independent pixel number. One thousand tests were performed to reduce the effect of random fluctuations.

The intent of the second simulation was to explore the effect of the number of cardiac cycles included in the image series. We constructed 12 image series with different β, corrupted with Gaussian IID noise, and then evaluated corresponding $V_c$. The quantitative relation between β and $V_c$ was studied by linear regression, and compared to the theoretical result [15].

In order to check the consistency of the results, another four cardiac MR cine series acquired in four volunteers using the same imaging parameters were processed through the same simulation and compared to the previous results.

### B. In-vivo real-time Cardiac MRI cine Images

16 free-breathing, dynamic cardiac MR cine series were acquired in six healthy volunteers using a real-time SSFP sequence [17, 18] using TSENSE [19] parallel imaging technique with acceleration factor four. Each cine series consisted of 256 frames spanning multiple cardiac and respiratory cycles, with each frame containing 192 × 144 pixels. The temporal resolution for each frame was approximately 60 msec. Since both the cardiac motion and the respiratory motion are quasi-periodic, the image data matrix should be represented by the sum of a low-rank signal matrix and a noise matrix. The number of relevant eigenimages was calculated using the RMT-fitting method for each images series.

## IV. Results

Both numerical simulations and in-vivo experiments demonstrated that the number of relevant eigenimages of a real-time cardiac cine can be determined by the RMT-fitting method. If we assume eigenvalues are uncorrelated random numbers, and use a significance level of 5% in the KS-test, no statistical significant deviation from the MP-law was observed throughout this study.

### A. Numerical Simulation

When the Gaussian IID noise was added, the critical noise variance $V_c$ was 66.0 × $\lambda_{min}$, see Fig. 1a. The mean pixel number from fitting was systematically smaller than the actual value, but the difference was less than 5% on average, see Fig. 1b.

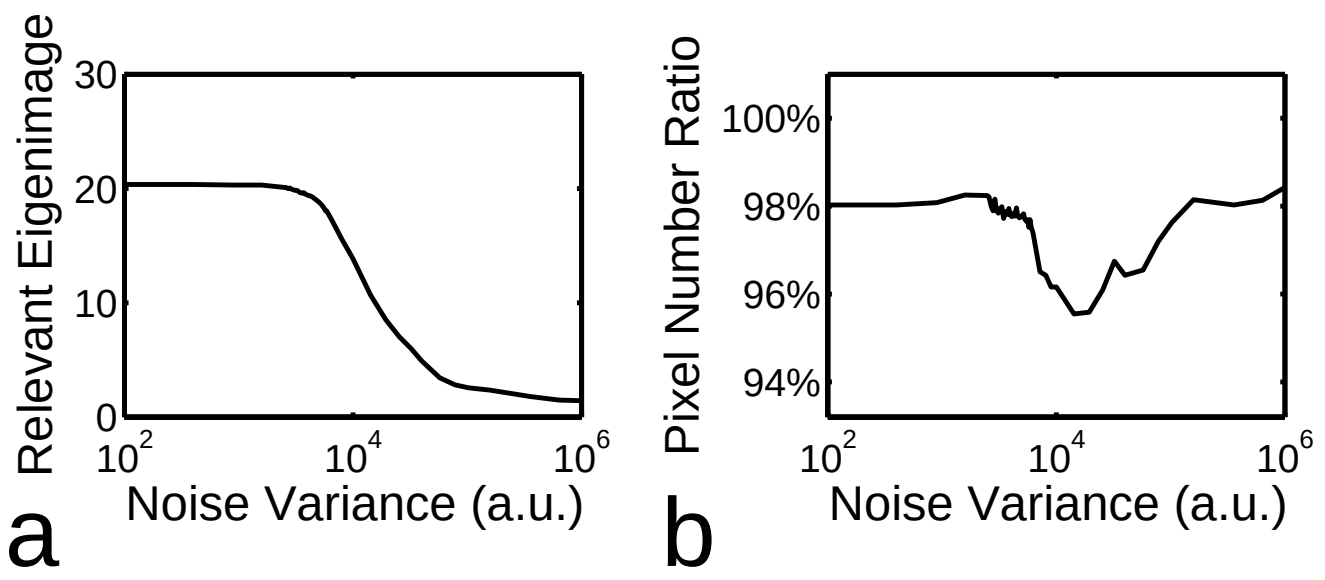

Fig. 1. The simulation results with Gaussian additive noise. (a) The number of relevant eigenimage vs. additive noise variance; (b) the fitted pixel number / actual vs. additive noise variance (the measurement SD < 5% for all data points).

When spatially correlated noise was added, the $V_c$ was 168.4 × $\lambda_{min}$ (measured before applying LPF). As expected, the number of independent pixels is close to 25% of the total, see Figure 2a. In the simulation of variable noise LPF cutoff, the RMT method resulted in the accurate number of independent pixels (difference < 5%), see Figure 2b.

In the spatially variant noise scenario, the $V_c$ is 60× $\lambda_{min}$ (measured after multiplying the template). The number of relevant eigenimages and the KS-test value are shown in Figure 3.

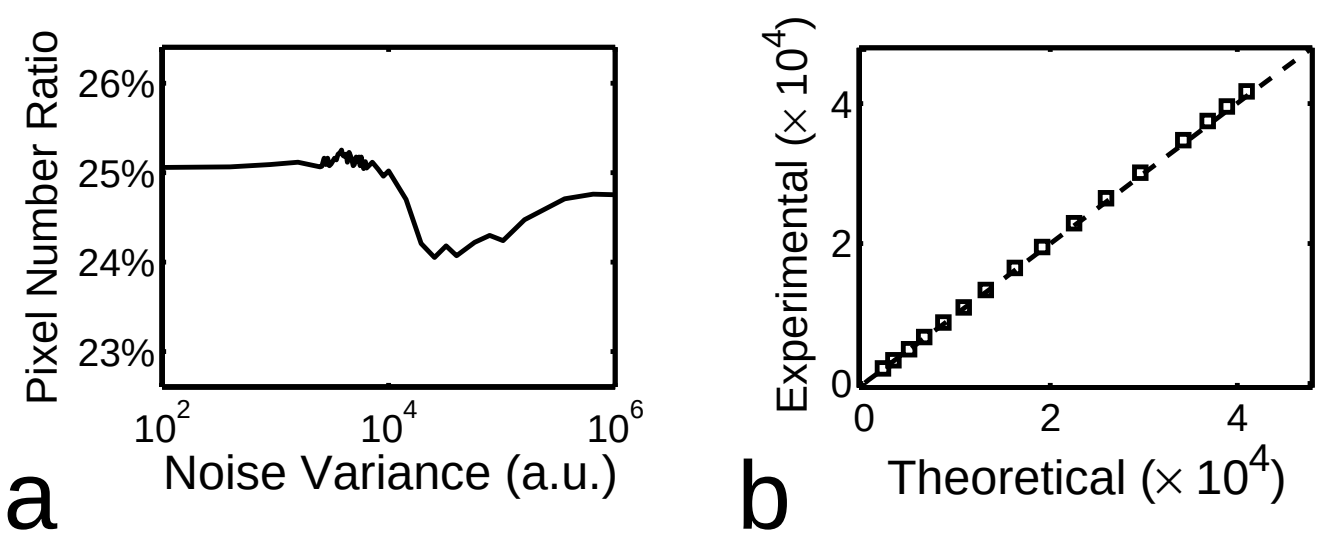

Fig. 2. The fitted independent pixel number. (a) The independent pixel ratio vs. additive noise variance, (b) the fitted vs. the actual independent pixel number at different LPF cutoffs. The dashed line indicates the function of Y = X.

Figure 4 shows the relation between $V_c$ and β. The theoretical relation is $(V_c / \lambda_{min})^2 = (N/M)\times\beta$ [15]. From our simulation, the relation is $(V_c / \lambda_{min})^2 = 0.42\times(N/M)\times\beta$, *i.e.* the RMT-fitting method can tolerate a slightly lower noise level.

Additionally, the mean noise variance can be assessed from the mean of the eigenvalues corresponding to the noise-only eigenimages as mentioned in the Theory section. The maximum noise variance measurement error in any single simulation test was small: 0.30% for the Gaussian noise case, 3.0% for the spatially variant noise case, and 1.2% for the spatially correlated noise case. Even when the number of relevant eigenimages is not accurate, the noise variance measurement is still accurate.

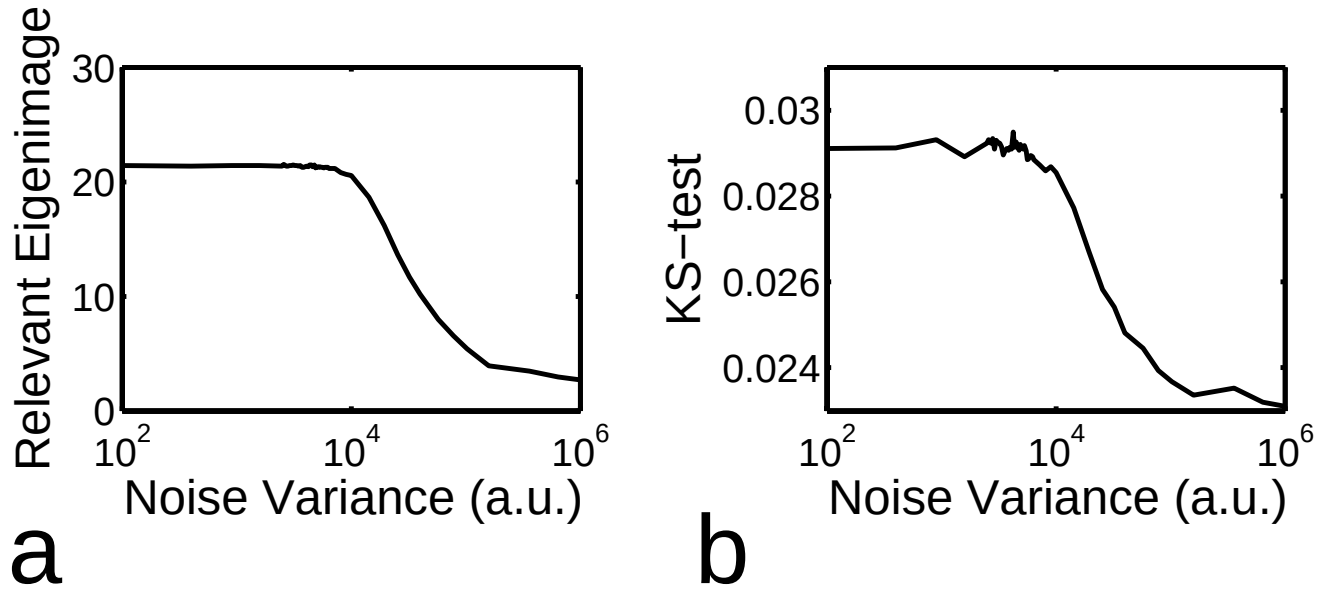

Fig. 3. (a) The number of relevant eigenimages vs. spatially correlated additive noise variance, (b) the KS-test value vs. spatially correlated additive noise

Numerical simulations using the additional four image series showed nearly identical results. We evaluated the relative fluctuations of critical noise variances (scaled by $\lambda_{min}$) for accurate number of relevant eigenimages measurements, pixel number measurements (scaled by the pixel number of an image) in the spatially correlated noise scenario, and noise variance measurements. They were less than 5%, 10%, and 4%, respectively.

*B. In-vivo real-time Cardiac MRI cine Images*

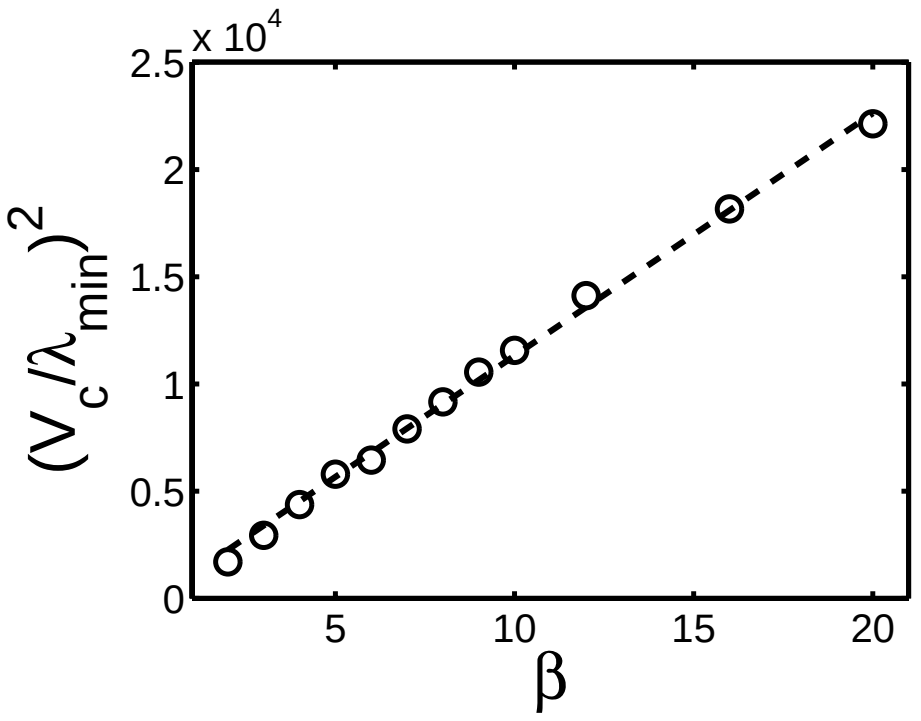

Fig. 4. The relation between critical noise variance $V_c$ and the repetition β. The linear regression (Y = cX) is represented by the dashed line.

The average number of relevant eigenimages selected by the RMT-fitting method was 90 ± 11. No clear spatially coherence structures were visible in the eigenimages determined to contain only noise. An example of the eigenvalue distribution and a fitted line according to Eq. (1) are shown in Figure 5a. Figure 5b shows the eigenvalue "scree" plot. There is no clear jump around the number of relevant eigenimages selected by the RMT-fitting method. Some unknown degree of spatial filtering is applied by the image reconstruction software to all images acquired on a commercial MRI system. Therefore, the pixel number generated by RMT-fitting method is smaller than image size. We may truncate all the noise-only eigenimages and reconstruct eigenfiltered images using relevant eigenimages. The mean noise variance in the filtered images is 90/256 ≈ 35% of the corresponding original images.

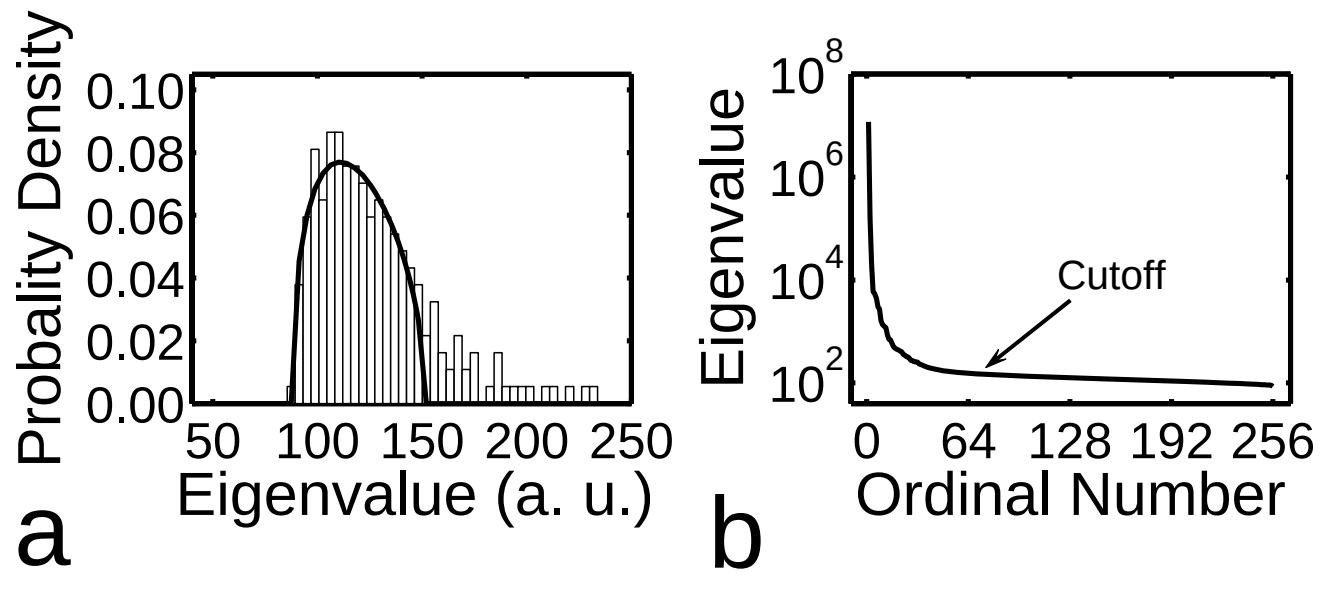

Fig. 5. The eigenvalue distribution and the cutoff of relevant eigenimage of an in-vivo real-time cardiac MRI cine image series. Two parameters determined by the KS-test are the following: the cutoff = 74 and the effective number of pixels = 10152, about 36.7% of the original. (a) The distribution of the last 238 eigenvalues and the prediction of the RMT (solid line). (b) The "scree" plot of all 256 eigenvalues.

All the results shown were acquired using the KS-test to determine GOF. When the same experiments were repeated using the CVM-test, the maximum relative difference between all results was within 5%. The CVM-test based RMT-fitting method has a smaller random fluctuation (~ 30% smaller in standard deviation).

## V. Discussion

The RMT-fitting method successfully identifies the number of relevant eigenimages in numerical simulations with three different noise distributions. Even with an additive noise variance over 60 times larger than $\lambda_{min}$, correct numbers of relevant eigenimages were found.

The quantitative relation between critical noise variance $V_c$ and repetition $\beta$ is also verified by our numerical simulation. This relation reveals the minimum number of repetitions needed for any signal mode to be distinguished from random noise using PCA. It indicates that all information modes in a noisy cardiac image series can be recovered using PCA as long as enough cardiac cycles are included, which guarantees that we can trade the scan time for SNR in real-time cardiac MR cine imaging. Comparing to the theoretical prediction [15], our method shows a lower noise tolerance (the noise standard deviation ≈ 80% of the theory). Two reasons may cause this result: first, there exist small deviations from MP-law, because the MP-law is the asymptotic eigenvalue distribution when $M, N \rightarrow \infty$ ; second, the sensitivity of the GOF we selected may be relatively low.

The in-vivo experiment shows that the MRT-fitting method can indentify the relevant eigenimages, and reveal the intrinsic low-rank structure of the corresponding data matrix corrupted by non-IID noise. As a by-product, a series of noise-only eigenimages can be identified. Therefore, the noise variance of the original image series can be accurately assessed. This is a new method to measure the temporal noise in dynamic images, which may be important to objectively evaluate the quality of dynamic images with non-IID noise.

Both the numerical simulation and the in-vivo imaging experiments demonstrate that the MP-law can fit the eigenvalue distribution reasonably well in the ($r$, $N$)-plane even when non-IID noise appears. Therefore, the MP-law is still a good approximation of the eigenvalue PDF when non-IID noise appears. When the noise spatial correlation is induced by the low-pass filtering, the number of independent variables no longer corresponds to the number of pixels. Therefore, it is understandable that the pixel number *N* should be allowed to vary when applying the MP-law in this scenario. A strict mathematical proof is warranted in the future.

## VI. CONCLUSION

The RMT-fitting method proposed in this paper provides a new solution to identify relevant eigenimages, and also a method to distinguish signal from noise. We believe it can be applied as a generic method to select relevant eigenmodes in PCA. Therefore, it is a very principled approach and will have impact on many important problems in related areas.

## ACKNOWLEDGMENT

Yu Ding thanks Prof. Heping He and Dr. Yan Su for helpful discussions.